\documentclass[reprint, amsmath,amssymb,aps]{revtex4-1}
\usepackage{graphicx}% Include figure files
\usepackage{dcolumn}% Align table columns on decimal point
\usepackage{bm}% bold math

\begin{document}

\title{Hybrid vertical-cavity laser with lateral emission into a silicon waveguide}

\author{Gyeong Cheol Park, Weiqi Xue, Alireza Taghizadeh, Elizaveta Semenova, Kresten Yvind, Jesper M{\o}rk, and Il-Sug Chung}
\email{ilch@fotonik.dtu.dk}

\affiliation{Department of Photonics Engineering (DTU Fotonik), Technical  University of Denmark, DK-2800 Kgs. Lyngby, Denmark}

\date{\today}

\begin{abstract}
% insert abstract here
We experimentally demonstrate an optically-pumped III-V/Si vertical-cavity laser with lateral emission into a silicon waveguide. This on-chip hybrid laser comprises a distributed Bragg reflector, a III-V active layer, and a high-contrast grating reflector, which simultaneously funnels light into the waveguide integrated with the laser. This laser has the advantages of long-wavelength vertical-cavity surface-emitting lasers, such as low threshold and high side-mode suppression ratio, while allowing integration with silicon photonic circuits, and is fabricated using CMOS-compatible processes. It has the potential for ultrahigh-speed operation beyond 100 Gbit/s and features a novel mechanism for transverse mode control.
\end{abstract}

\maketitle

\section{Introduction}
\label{sec:intro}

\begin{figure}[b]% fig. 1
  \centering
  \includegraphics[angle=0, width=0.7\linewidth]{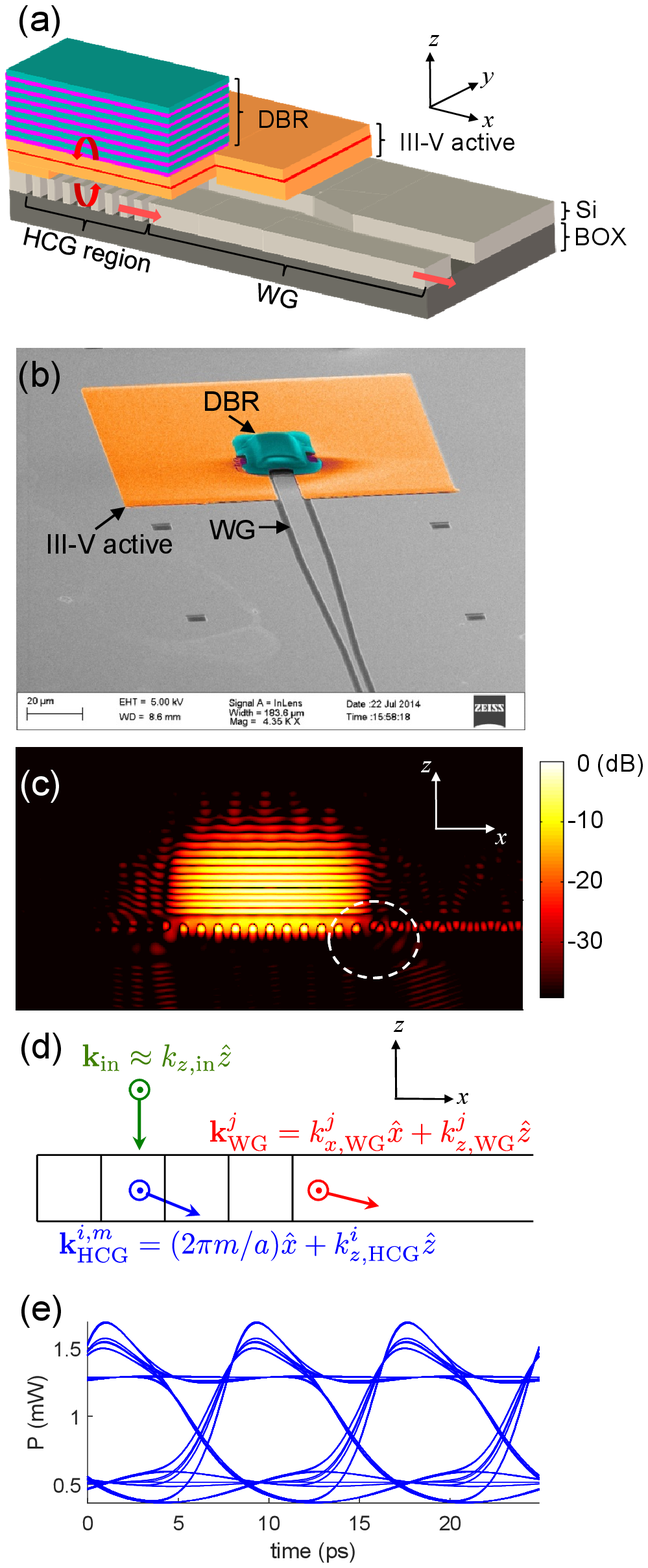}
  \caption{(a) Schematic cross-sectional view of the hVCL. (b) Scanning electron microscope image of the fabricated hVCL. (c) Fundamental mode profile of the hVCL. (d) $H_y$ fields and wavevectors of modes within the white dotted circle in Fig. 1(c). $\mathbf{k}_\mathrm{HCG}^{i,m}$: wavevector of the $m$-th harmonic of the $i$-th HCG eigenmode. $\mathbf{k}_\mathrm{WG}^{j}$: wavevector of the $j$-th waveguide mode. $a$: grating period. (e) Numerical simulation of the eye-diagram at 120 Gbit/s with a PRBS of $2^7-1$ and a current modulated between 6 and 14 mA.}
\label{Fig.1}
\end{figure}

A laser that can be integrated onto a silicon (Si) wafer and act as an on-chip light source for Si photonics applications, e.g.~chip-level optical interconnects, is in much demand and requires innovative breakthroughs in laser diode technology and design. The ideal on-chip laser source needs to meet several criteria \cite{Miller_2009}: (1) The energy consumption in generating an optical bit (energy/bit) should be no more than 10 fJ/bit for on-chip interconnects and around 100 fJ/bit for off-chip interconnects. (2) The bit error rate (BER) for the received signal should preferably be less than $10^{-12}$, putting strict requirements on laser output power and noise. At 10 Gb/s, the laser output power thus needs to  be well above 10 $\mu$W \cite{Agrawal_4ed}, with higher bit rates requiring higher output powers. 

Several interesting on-chip laser structures have already been reported, including the Si evanescent laser (SEL) \cite{Park_2005, Zhang_2014}, the lambda-scale embedded active region photonic-crystal (LEAP) laser \cite{Matsuo_2010, Takeda_2013}, and the Ge laser \cite{Camacho-Aguilera_2012}. While promising, it appears difficult to meet both criteria simultaneously. They have a challenge in either of the two criteria. The vertical-cavity surface-emitting laser (VCSEL) has been the most successful light source for short-reach interconnects (a few meters to less than 1 km). Since this micro laser can provide an energy/bit as low as 69 fJ/bit \cite{Hofmann_2012} as well as safely satisfying the BER requirement, it can also be an attractive on-chip light source if the emission can be directed efficiently into a Si waveguide, as discussed in \cite{Takeda_2013}. 

In this letter, we report what we believe to be the  first experimental demonstration of the hybrid vertical cavity laser (hVCL) with in-plane output to a Si waveguide. Furthermore, as an important technological achievement, the laser was fabricated using direct wafer bonding and CMOS (complementary metal-oxide-semiconductor) compatible processes \cite{Bowers_2009}. 
The first idea of the hVCL was proposed in \cite{Chung_2010}, based on numerical simulations. Sciancalepore et al. have experimentally shown that light can be coupled from a vertical cavity to an in-plane Si waveguide, but without evidence of lasing \cite{Sciancalepore_2012}. 

\section{Device structure}
As shown in Fig. 1(a), the hVCL structure comprises a 7-pair TiO$_2$/SiO$_2$ distributed Bragg reflector (DBR), a III-V active layer, a high-index-contrast grating (HCG) reflector, and a Si waveguide abutted to the HCG region. The III-V active layer is $1\lambda$ thick and includes 7 InGaAlAs/InGaAlAs strained quantum wells (QWs). Both the HCG and the Si waveguide are formed in the Si layer of a Si-on-insulator (SOI) wafer \cite{Chung_2010}. 

As the red arrows in Fig. 1(a) illustrate, light is amplified vertically between the DBR and the HCG reflector, and is emitted through the Si waveguide. The HCG is designed to reflect most of a vertically incident mode, e.g., 99.5 \% and route a small fraction, e.g., 0.3 \% into the Si waveguide. This routing of 0.3 \% corresponds to an output-coupling efficiency of about 40 \%, which is defined as the fraction of the in-plane output power over the sum of all optical losses including mirror, scattering, and absorption losses. The mode profile in Fig. 1(c) which is obtained by finite-difference time-domain (FDTD) simulations, shows that light is emitted from the vertical cavity mode laterally into the Si waveguide. As illustrated in Fig. 1(d), the vertically incident mode with wavevector, $\mathbf{k}_\mathrm{in}$  excites the eigenmodes of the HCG with wavevectors, $\mathbf{k}^{i,m}_\mathrm{HCG}$. Since the $\mathbf{k}^{i,m}_\mathrm{HCG}$ has harmonic components along the $x$ direction, the HCG eigenmodes can couple to the Si waveguide mode with wavevector, $\mathbf{k}^j_\mathrm{WG}$. This mode conversion from the HCG modes to the Si waveguide mode occurs in the connecting part denoted by a white dotted circle in Fig. 1(c). 

As shown in Fig. 1(c), there is no field penetration below the HCG, which results in a confinement factor as high as 11.8 \%. This high confinement factor leads to very high intrinsic modulation speed of 120 Gbit/s, as shown in Fig. 1(e). The numerical eye-diagram simulation uses rate equations, including gain suppression and cavity parameters extracted from FDTD simulations. This high speed potential of the hVCL is promising for high speed chip-level optical interconnects.

\section{Fabrication and Characterization}
\label{sec:fabrication and characterization}

The sample shown in Fig. 1(b), was fabricated by using CMOS-compatible processes: 1) electron-beam lithography for forming a HCG region and a waveguide on a SOI wafer, 2) a direct wafer bonding process for hybridizing the III-V active layer onto the SOI wafer \cite{Bowers_2009}, and 3) a lift-off process for forming a DBR. In addition, selective wet etching was conducted to form an air gap between the III-V active layer and the HCG. For characterisation, the fabricated hVCL sample was cleaved so as to expose the Si waveguide end to the air.

The fabricated hVCL sample was optically pumped by using a 980-nm diode laser in pulsed operation mode with a 2-\% duty cycle ($f_\mathrm{pump}$) and a 5-MHz repetition rate. The hVCL sample was un-cooled. Pumping at 1300 nm  would avoid light absorption in the Si layers and subsequent heating, but was not possible in the present samples since the DBR stop band extends to 1300 nm. This is, however, merely a design issue.  Due to heat generation,  pulsed operation was required. The pumping laser beam was focused onto the top DBR using alignment optics that include a 50$\times$ near-infrared objective lens. The white dotted circle in Fig. 2(a) denotes the position of the pumping beam. The measured power transmittance of the alignment optics, $T_\mathrm{a}$ is 21 \%. The pumping power is monitored before the alignment optics. The pumping beam profile on the sample surface and the vertical output profile were imaged by using the alignment optics, an InGaAs camera, and a long-wavelength-pass filter, as shown in Figs. 2(b) and 2(c), respectively. The $e^{-2}$ diameters for the pumping beam, $d_\mathrm{pump}$, and the vertical output, i.e., transverse mode, $d_\mathrm{mode}$ are 5.84 $\mu$m and 5.99 $\mu$m, respectively. The in-plane output was collected by a multi-mode fiber at the end facet of the Si waveguide. The beam profile of the in-plane output was characterised by using a lensed fiber, as shown in the inset of Fig. 2(e). 
\section{Results and discussion}
\label{ssec:results and discussion}

\begin{figure}
  \centering
  \includegraphics[angle=0, width=0.6\linewidth]{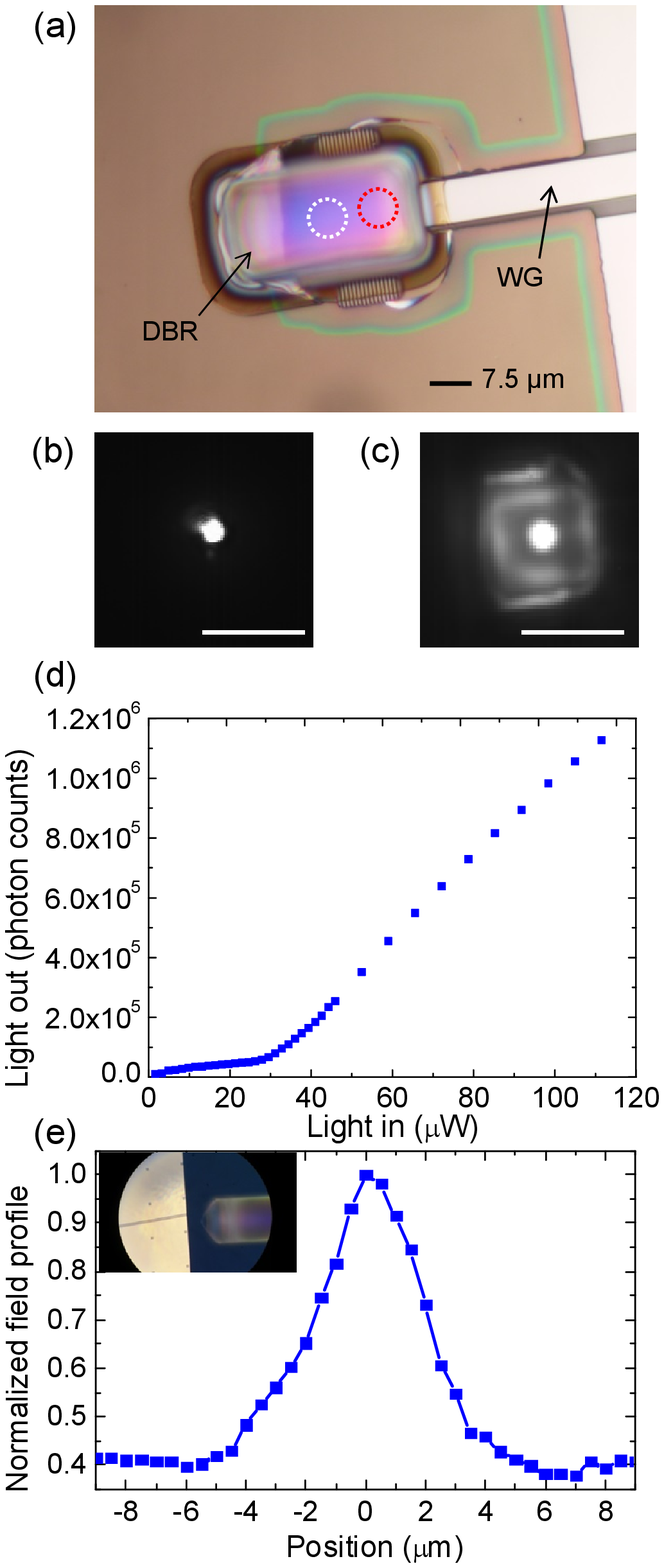}
  \caption{(a) Microscope image of the hVCL sample seen from the top. Near-field images of (b) the pumping beam and (c) the vertically-emitted lasing output. The scale bar is 17.5 $\mu$m. (d) In-plane light output from the Si waveguide as a function of absorbed input light. (e) Intensity profile of the in-plane output from the waveguide. The position '0' corresponds to the waveguide position.}
  \label{Fig.2}
\end{figure}

Figure 2(d) shows the measured light-out, $L_\mathrm{out}$ versus light-in, $L_\mathrm{in}$ (LL) curve. The light-in represents the power absorbed in the QWs and separate confinement heterostructure (SCH) layers, and is estimated as $L_\mathrm{in}=P_\mathrm{m} \times T_\mathrm{a} \times (d_\mathrm{mode}/d_\mathrm{pump})^2 \times A$, where $P_\mathrm{m}$ is the monitored power incident to the alignment optics and $A$ is the absorption efficiency. The absorption efficiency measures the fraction of the absorbed power in the QWs and SCH layers over the power incident to the hVCL and is estimated by FDTD simulations to be 30 \%. The LL curve clearly shows that lasing starts at a threshold input power, $P_\mathrm{th}$ of about 29 $\mu$W. The corresponding threshold current, $I_\mathrm{th}$ for the continuous wave (CW) case can be estimated as 
\begin{equation}
I_\mathrm{th}= 29\mu\mathrm{W}/E_{980}\times q/f_\mathrm{pump}\sim 1.1~\mathrm{mA},
\end{equation}
where $E_{980}$ is the photon energy at 980 nm and $q$ is the electron charge. The estimated $I_\mathrm{th}$ is similar as that of long-wavelength VCSELs with a similar mode size \cite{Muller_2011}. 

The output power in the Si waveguide can be increased to the mW-level. Numerical studies based on FDTD simulations indicate that the output efficiency from the vertical cavity to the Si waveguide is maximised when the tail of the transverse mode overlaps with the starting position of the Si waveguide \cite{Chung_2010}. The red dotted circle in Fig. 2(a) denotes the  pumping position that gives this optimal overlap. However, the laser output power is maximised when the central part of the DBR, indicated by the white circle, is pumped. This is attributed to non-uniformity in the DBR thickness. In Fig. 2(a), the color difference on the DBR surface between the white and red circle positions reflects that the DBR thickness within the red circle is smaller than within the while circle. This implies that the local reflectivity is smaller within the red circle, increasing the laser threshold: thus we could not reach lasing with the available pumping power. Since the optimal position for pumping is displaced by about 10 $\mu$m from the optimal position for lateral waveguide output, the output efficiency from the vertical cavity to the Si waveguide becomes much lower than the designed value. The thickness non-uniformity problem caused by the lift-off process can be solved by using an etch-back process to define a DBR mesa and/or the hVCL design can be changed so that the optimal pumping position be located in the center of a DBR. Based on numerical simulations, these changes are expected to increase the output power to a few mW in the Si waveguide \cite{Chung_2010}. 

The output light is well confined within the Si waveguide. As shown in Fig. 2(e), the beam profile is clearly centered at the Si waveguide position. The full width at half maximum (FWHM) of the beam profile is  $\sim$5 $\mu$m which is larger than the waveguide end width of 1.5 $\mu$m. This difference is attributed to the divergence of the output beam from the small-aperture waveguide. 

\begin{figure}
  \centering
  \includegraphics*[angle=0, width=0.7\linewidth]{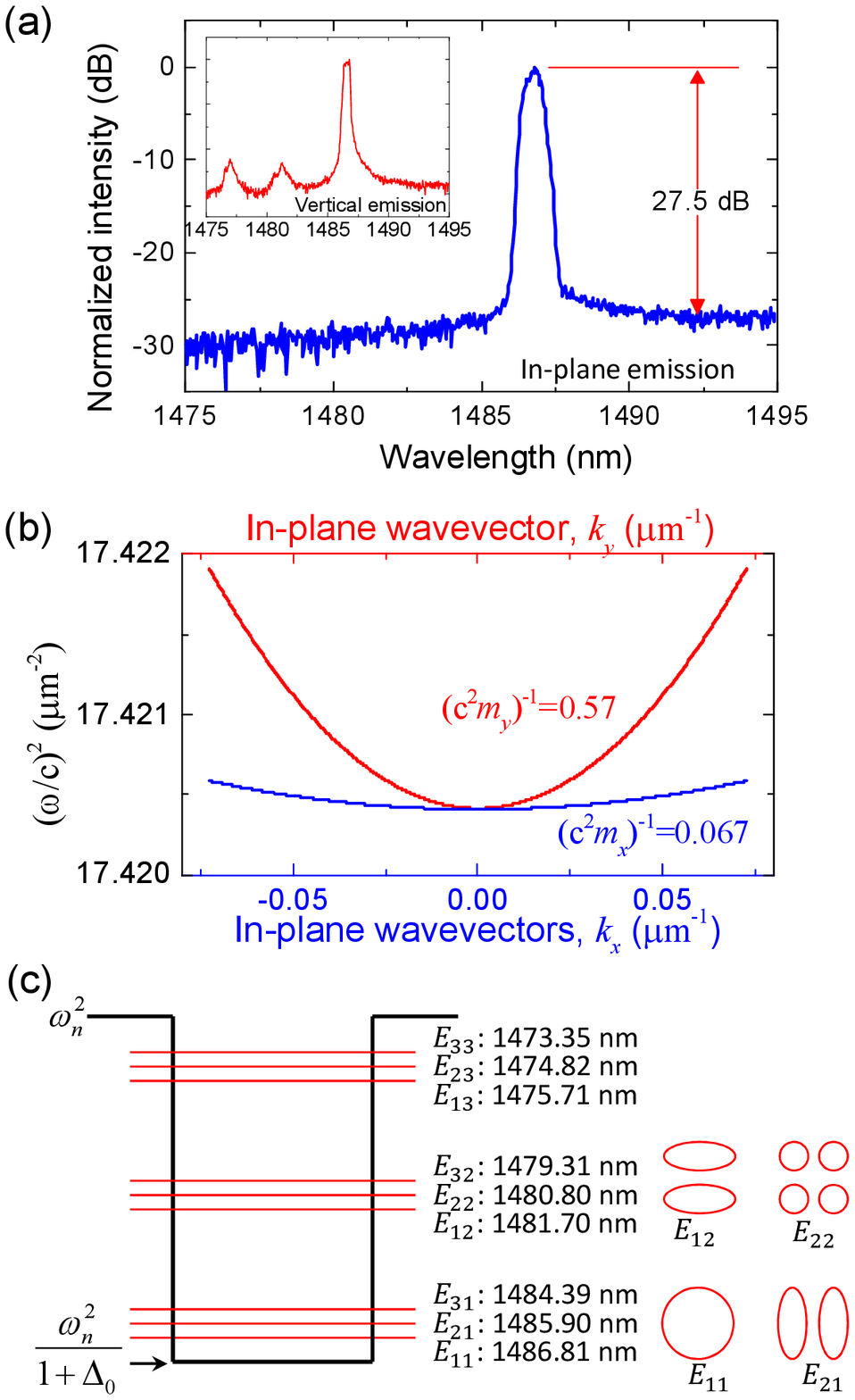}
  \caption{(a) Measured in-plane output spectrum at $4P_\mathrm{th}$ and the vertical emission spectrum at $P_\mathrm{th}$. (b) Vertical-cavity resonance angular-frequency square versus in-plane wavevectors. (c) Band-edge envelope of a photonic quantum well, confined states $E_{ij}$ with wavelengths, and corresponding transverse mode profiles.}
  \label{Fig.3}
\end{figure}

Figure 3(a) shows the in-plane emission spectrum measured from the Si waveguide at a pumping power of $4P_\mathrm{th}$ as well as the vertical emission spectrum measured around $P_\mathrm{th}$. Around the threshold, there are three peaks with a spacing of 4.5 nm to 5 nm. At $4P_\mathrm{th}$, the longest wavelength peak becomes dominant with a side-mode suppression ratio (SMSR) of 27.5 dB. It is noteworthy that the spectral linewidth (FWHM) of the lasing peak is 0.66 nm, which is large considering that the pumping power is well above the threshold. In fact, it appears to comprise about 3 sub peaks. If there are three sub peaks, the sub mode spacing is about 0.2 nm to 0.3 nm. These seemingly groups of sub-peaks are firstly observed for the VCL with a HCG. 

We find that the appearance of sub peaks can be attributed to anisotropic transverse quantum confinement due to the HCG. The dependence of a vertical resonance angular frequency, $\omega$ on transverse wavevectors, $k_x$ and $k_y$ is largely influenced by the transverse dispersion of the HCG constituting the vertical cavity \cite{Boutami_2008}. As shown in Fig. 3(b), our vertical cavity has a highly anisotropic transverse dispersion: the band curvature in the direction perpendicular to grating bars, $\partial^2 \omega^2 / \partial k_x^2$ ($\equiv 1/m_x$) is 8.5 times smaller than that in the parallel direction, $\partial^2 \omega^2 / \partial k_y^2$ ($\equiv 1/m_y$). In our laser sample, the absorption loss in the QWs outside the pumping region determines the transverse mode size, which is confirmed by FDTD simulations. The exponential decay of the transverse mode in the lossy region is equivalent to the decay in the barrier region of a quantum well. Thus, we may model the band-edge frequencies of the pumping and lossy regions as shown in Fig. 3(c). Then, angular frequency $\omega_{i,j}$ for confined transverse state $E_{i,j}$ can be derived by using the envelope approximation for photonic crystal heterostructures \cite{CL_2002}:

\begin{equation}
\omega^2_{i,j} = \frac{1}{1+\Delta_0}\left[\omega_n^2 + \frac{(\alpha_i~ i\pi)^2}{2m_x L_x^2} + \frac{(\beta_j j\pi)^2}{2m_y L_y^2}\right],
\end{equation}  

where $i$ and $j$ are quantum numbers for $x$ and $y$ directions, respectively, and $\alpha_i$ and $\beta_j$ are rational factors, due to the finite barrier height. Angular frequency $\omega_n$ and $\Delta_0$ determine the cut-off frequency and the barrier height, respectively. Quantum well widths, $L_x$ and $L_y$ are assumed to be $d_\mathrm{pump}$. The calculated wavelengths of confined states $E_{i,j}$ listed in Fig. 3(c), reproduce the grouping of modes in Fig. 3(a), which is a good agreement, considering the simplicity of this model. Several transverse mode profiles are listed in Fig. 3(c). The lasing peak should be a mixture of $\{E^x_{i1}\}$. This result shows that the anisotropic dispersion in a HCG-based vertical cavity has huge influence on transverse mode wavelength and profile, which needs to be carefully considered when designing HCG-based vertical cavities. 

In the electrically-pumped version of hVCLs, HCG heterostructure will be used for both $x$- and $y$-direction. Then, not only the band curvature, $m_i$ but also the barrier height, $\Delta_0$ can be controlled separately for $x$ and $y$ directions. This provides rich design flexibilities for transverse mode control. Details based on 3D simulations will be reported elsewhere.

\section{conclusion}
\label{ssec:Conclusion}

We have experimentally demonstrated a hybrid vertical-cavity laser (hVCL) with lateral emission into a Si waveguide, using CMOS-compatible processes. This on-chip laser has a threshold comparable to VCSELs, and shows good suppression of higher order modes. The laser has the potential to be modulated at bitrates larger than 100 Gbit/s due to a very high confinement factor,  and is capable of novel transverse mode control. These properties make the hVCL a promising candidate as a low-cost high-performance light source in chip-level optical systems. \newline

\begin{acknowledgements}
	The authors gratefully acknowledge supports by the Danish Research Council through the FTP project (Grant No. 11-106620) as well as the Villum Fonden via the NATEC Centre of Excellence.
\end{acknowledgements}

\end{document}